\documentclass[aps,prd,twocolumn,superscriptaddress,nofootinbib,10pt]{revtex4-1}
\usepackage[paperwidth=21cm,paperheight=29.7cm,top=2.54cm,bottom=2.54cm,left=2cm,right=2cm]{geometry}
\usepackage[colorlinks,linkcolor=blue,anchorcolor=blue,citecolor=blue,urlcolor=blue]{hyperref}
\usepackage{graphicx,subfigure}
\usepackage[figuresright]{rotating}
\usepackage{amsmath,amsfonts,amssymb,bm}
\usepackage{array,enumitem,multirow}
\usepackage{acronym}
\usepackage{makecell}
\newcommand{\be}{\begin{equation}}
\newcommand{\ee}{\end{equation}}
\newcommand{\bea}{\begin{eqnarray}}
\newcommand{\eea}{\end{eqnarray}}

\newacro{GR}{general relativity}
\newacro{GW}{gravitational wave}
\newacro{MG}{modified gravity theory}
\newacro{BH}{Black hole}
\newacro{PN}{post-Newtonion}
\newacro{ppE}{parameterized post-Einsteinian}
\newacro{GCB}{galactic ultra-compact binary}
\newacro{SBHB}{stellar-mass black hole binary}
\newacro{MBHB}{massive black hole binary}
\newacro{BHB}{black hole binary}
\newacro{IMBHB}{intermediate-mass black hole binary}
\newacro{EMRI}{extreme mass ratio inspiral}
\newacro{IMRI}{intermediate mass ratio inspiral}
\newacro{SGWB}{stochastic gravitational wave background}
\newacro{MECO}{minimal energy circular orbit}
\newacro{FAR}{false alarm rate}
\newacro{CE}{Cosmic Explorer}
\newacro{ET}{Einstein Telescope}
\newacro{LISA}{Laser Interferometer Space Antenna}
\newacro{EdGB}{Einstein-dilaton Gauss-Bonnet}
\newacro{dCS}{dynamic Chern-Simons}
\newacro{SNR}{signal-to-noise ratio}
\newacro{FIM}{Fisher Information Matrix}
\newacro{ISCO}{innermost stable circular orbit}
\newacro{NSBH}{neutron star-black hole binary}
\newacro{MCMC}{Markov Chain Monte Carlo}
\newacro{QNM}{quasi-nomral mode}

\begin{document}

\title{ Boundary conditions of general black hole perturbations}

\author{Wei Xiong}
 \email{202210187053@mail.scut.edu.cn}%
\author{Peng-Cheng Li}%
 \email{pchli2021@scut.edu.cn, corresponding author}
\affiliation{%
School of Physics and Optoelectronics, South China University of Technology, Guangzhou 510641, People’s Republic of China.
}%

\date{\today}

\begin{abstract}	
Recently, significant progress has been made in the study of black hole (BH) perturbations, both within the framework of general modified gravity theories and in complex environments. However, a well-established conclusion regarding the boundary conditions of the perturbed fields remains elusive. In this paper, we investigate the boundary conditions for a general perturbation at spatial infinity and the event horizon of a BH described by a generic metric that is stationary, axisymmetric, asymptotically flat, and respects the condition of circularity. Our analysis is independent of any specific BH model or the nature of the perturbed field. In particular, by extending the formulation introduced by Teukolsky and utilizing purely geometric principles, we derive a universal expression for the boundary condition at the horizon. This expression is elegantly formulated in terms of the physical quantities at the event horizon, specifically the BH's surface gravity and angular velocity. The results presented in this work may provide valuable insights for the calculation of quasinormal modes and the gravitational waves generated by extreme-mass-ratio inspirals, extending beyond the standard Kerr case.
\end{abstract}

\maketitle


\section{Introduction}
\label{section1}

The black hole (BH) perturbation theory plays a crucial role in the gravitational wave physics \cite{Kokkotas:1999bd,Berti:2009kk,Konoplya:2011qq}. It can be applied to the calculation of quasinormal modes (QNMs) for the ringdown phase of binary mergers \cite{Berti:2005ys,Berti:2018vdi}, and the waveform generation of extreme-mass-ratio inspirals \cite{Amaro-Seoane:2012lgq}. 
 Moreover, the BH perturbation theory is instrumental in testing the stability of BH solutions in numerous gravity models \cite{Herdeiro:2015waa,Doneva:2022ewd,Dias:2015nua}. Several significant BH models have been found to be unstable, such as the scalar hairy Kerr BH \cite{Ganchev:2017uuo}. In general, the BH perturbation theory involves perturbing a fixed background (the metric and matter fields) and retaining only the leading order of the perturbed fields in the equations of motion. The resulting perturbation equations become a set of linear differential equations, which is more reduced than the full equations. By imposing specific conditions at boundaries (e.g., the ingoing wave at the event horizon and the outgoing wave at the radial infinity for the asymptotically flat BHs) and the symmetry of the spacetime, the perturbation equations can then be solved. 

The perturbation equations of static spherically symmetric BHs can typically be decoupled into radial and angular ordinary differential equations \cite{Regge:1957td,Zerilli:1970wzz}. Similarly, for Kerr BHs and spinning C metric, which are of Petrov type D, perturbations are described by a single Teukolsky equation and variables can be separated within the Newman-Penrose (NP) formalism \cite{Teukolsky:1972my,Teukolsky:1973ha,Bini:2008mzd}. The radial equations can then be transformed into a single second-order equation in the Schr\"odinger-like form \cite{Nagar:2005ea,Guo:2024bqe} (however, see \cite{Langlois:2021aji,Langlois:2022eta,Roussille:2023sdr} for an alternative treatment). The condition at each boundary can be obtained straightforwardly. For more complex situations, such as Kerr-Newman (KN) BHs \cite{Mark:2014aja,Dias:2015wqa}, the Einstein-scalar-Gauss-Bonnet (EsGB) models \cite{Blazquez-Salcedo:2018jnn,Blazquez-Salcedo:2020caw,Blazquez-Salcedo:2020rhf,Blazquez-Salcedo:2017txk,Antoniou:2024gdf,Chung:2024ira,Antoniou:2024hlf,Chung:2024vaf,Blazquez-Salcedo:2024oek,Xiong:2024urw,Khoo:2024agm,Blazquez-Salcedo:2024dur} and Einstein-Maxwell-scalar (EMs) models \cite{LuisBlazquez-Salcedo:2020rqp,Myung:2018jvi,Blazquez-Salcedo:2019nwd}, the gravitational perturbations are coupled with the perturbations of other fields and the separation  of variables becomes infeasible. 

Recently, researches have made significant advance in numerically calculating the QNMs  for rotating BHs  \cite{Dias:2015nua,Chung:2023zdq,Chung:2023wkd,Blazquez-Salcedo:2023hwg,Xiong:2024urw,Blazquez-Salcedo:2024oek}. 
However, theses studies primarily focus on solving the perturbation equations for rotating BHs, and the detailed discussion on how to formulate the corresponding boundary conditions in a general model is almost absent. To the best of our knowledge, the only exception is \cite{Chung:2024vaf}, which offers a detailed analysis of the boundary conditions for perturbations of rotating BHs, employing a metric ansatz designed to parametrize small deviations from the Kerr metric \cite{Cano:2019ore}.  Clearly, transforming a set of coupled equations into a single Schr\"odinger-like equation is challenging. This process becomes increasingly impractical when dealing with models that include extremely complex perturbation equations, such as the EsGB models \cite{Blazquez-Salcedo:2020caw}. In such cases, the coefficients of these equations often need to be determined numerically, which further complicates the boundary analysis \cite{Blazquez-Salcedo:2024oek}.  

In this paper, we aim to analyze the boundary conditions for neutral perturbations of a general spinning BH that satisfies the following properties: it is axisymmetric, stationary, asymptotically flat, and respects circularity \cite{Papapetrou:1966zz, Delaporte:2022acp}. The ``neutral" means that the perturbation field does not interact with the background electromagnetic field. Although we aim to formulate the boundary analysis for the metric perturbation in complex system, we will not specify any explicit perturbed fields or models. We begin by reviewing the derivation of the boundary condition at horizon for the perturbations of Kerr BHs within the NP formalism. Through the formulation from Teukolsky, a boundary condition expressed in terms of solely the physical quantities at the horizon can be obtained \cite{Teukolsky:1973ha}. We argue that both the formulation of Teukolsky and the expression of the boundary condition can be generalized to the general cases. We then utilize a metric ansatz commonly used in the numerical construction of the stationary axisymmetric BH solution for numerous models \cite{Guo:2023mda,Xiong:2023bpl,Cunha:2019dwb,Herdeiro:2020wei,Herdeiro:2014goa,Herdeiro:2015gia,Herdeiro:2015tia,Herdeiro:2016tmi,Santos:2020pmh}, and impose two requirements on the perturbations. Firstly, the separation of variables for the perturbation function is based on the symmetry of the spacetime. The second is that the perturbation obeys the formulation from Teukolsky, namely the perturbation seen by a physically well-behaved observer at the horizon is nonspecial and ingoing. As a result, we derive the boundary condition at the event horizon for general cases, and find that it is consistent with the boundary condition for Kerr BHs. For completeness, we also derive the boundary condition for perturbations at spatial infinity. To validate the reasonability of our results, we analyze several specific models, including BHs in GR alternative to the Kerr BH, as well as BHs in various gravity theories where the boundary conditions for perturbations are either known or straightforward to derive.

We organize this paper as follows.  In Sect.\ref{section2}, we briefly review the boundary conditions for the Kerr BH, and extend the formulation to the general metric ansatz. In Sect.\ref{section3}, the verification of our conclusion for other type D BHs in GR, static hairy BHs and a test scalar field perturbation around a general rotating BH is presented,  respectively. We discuss in Sect.\ref{section4}. The geometric units $G=c=1$ are maintained throughout this paper.  
\section{Boundary conditions for general perturbations}
\label{section2}

\subsection{Teukolsky formulation}

In this subsection, we review the boundary condition at the event horizon for various perturbations in the background of Kerr BHs. 
The Kerr metric in terms of the well-known Boyer-Lindquist (BL) coordinates is expressed as \cite{Boyer:1966qh}
\begin{eqnarray}
	ds^{2} &=& -\frac{\Delta}{\Sigma} (dt-a\sin^{2}\theta d\varphi)^{2} + \frac{\Sigma}{\Delta} dr^{2} + \Sigma d\theta^2 \nonumber \\
	&&+ \frac{\sin^{2}\theta}{\Sigma} \left[ adt-(r^{2}+a^{2}) d\varphi \right]^{2}, \label{eq:Kerrmetric}
\end{eqnarray}
where $\Sigma \equiv r^{2} + a^{2}\cos^{2}\theta$ and $\Delta \equiv r^{2} -2 M r +a^{2}$. Here $M$ is the BH mass and $a$ is the BH spin, both of which are conserved charges evaluated at the asymptotic infinity. The general perturbed fields $\Psi$ of  the Kerr BH can be described by a single master equation, known as the Teukolsky equation \cite{Teukolsky:1972my}
\begin{eqnarray}
	&& -\left[ \frac{(r^{2}+a^{2})^{2}}{\Delta} - a^{2} \sin^{2}\theta \right] \frac{\partial^{2} \Psi }{\partial t^{2}} - \frac{4Mar}{\Delta} \frac{\partial^2 \Psi}{\partial t \partial \varphi} \nonumber \\
	&&- \left[ \frac{a^{2}}{\Delta} - \frac{1}{\sin^{2}\theta}  \right] \frac{\partial^{2} \Psi}{ \partial \varphi^{2}} + \Delta^{-s} \frac{\partial}{\partial r} \left( \Delta^{s+1} \frac{\partial\Psi}{\partial r} \right) \nonumber \\
	&&  + \frac{1}{\sin \theta} \frac{\partial}{\partial\theta} \left( \sin \theta \frac{\partial \Psi}{ \partial \theta} \right) + 2s \left[ \frac{a(r-M)}{\Delta} + \frac{i  \cos \theta}{\sin^{2} \theta} \right] \frac{\partial\Psi}{\partial\varphi}  \nonumber \\
	&& +2s \left[ \frac{M(r^{2}-a^{2})}{\Delta} - r - i a \cos \theta \right] \frac{\partial \Psi}{\partial t} \nonumber \\
	&&  - (s^{2} \cot^{2}\theta -s) \Psi  =4\pi \Sigma T_{(s)} ,
	\label{eq:Teukolsky}
\end{eqnarray}
The parameter $s$ represents the spin weight of the perturbed fields and $ T_{(s)} $ is the source term. For example,  $s=\pm 2$ denotes the gravitational perturbations. Under the separation of variables $\Psi = e^{-i\omega t + i m \varphi}  R(r) S(\theta)$, the master equation for the vacuum case ($T=0$) can be separated as the radial part
\begin{eqnarray}
	&&\Delta^{-s} \frac{d}{dr} \left( \Delta^{s+1} \frac{dR(r)}{dr} \right) + \nonumber \\
	&&\left( \frac{K^{2} - 2is(r-M)K}{ \Delta} + 4is\omega r -\lambda \right) R(r)= 0,
	\label{eq:radialTeukolsky}
\end{eqnarray}
and the angular part
\begin{eqnarray}
	&&\frac{1}{\sin \theta} \frac{d}{d\theta} \left( \sin\theta \frac{dS}{d\theta} \right) +   \nonumber \\
	&&\left(a^{2}\omega^{2}\cos^{2}\theta - 2a\omega s\cos \theta   -\frac{(m+s \cos\theta)^{2}}{\sin^{2} \theta}  +s +A \right) \nonumber \\
	&&\quad\times S(\theta) = 0,
	\label{eq:angularTeukolsky}
\end{eqnarray}
with $K \equiv (r^{2} + a^{2}) \omega -am$ and $\lambda \equiv A + a^{2} \omega^{2} -2am\omega$. Here $A$ is a separation constant, which can be viewed as the Sturm-Liouville eigenvalue problem for the angular equation (\ref{eq:angularTeukolsky}) together with the regularity boundary conditions at $\theta=0,\pi$. The eigenfunction $S(\theta)$ corresponding to each $A$ is known as the spin-weighted spheroidal harmonic function. By convention, before discussing the boundary condition of the radial equation (\ref{eq:radialTeukolsky}) at the horizon, one can perform a transformation
\begin{eqnarray}
	Y &=& \Delta^{\frac{s}{2}} (r^{2}+a^{2})^{\frac{1}{2}} R, \nonumber \\  
	r_{*} &=& \int\frac{r^{2}+a^{2}}{\Delta} dr, \nonumber \\ 
	 &=& \frac{r_{+}+r_{-}}{r_{+}-r_{-}} \left[ r_{+} \ln(r-r_{+}) - r_{-} \ln(r-r_{-}) \right] \nonumber \\
	&& +r  + c_{*},
	\label{eq:transformationTeukolsky}
\end{eqnarray}
where $c_{*}$ is an arbitrary constant from the integration and can be set to $0$, and $r_{\pm} \equiv M \pm \sqrt{M^{2} -a^{2}}$ represent the outer (event) horizon and inner horizon for the Kerr BH. The new coordinate $r_{*}$ defined in (\ref{eq:transformationTeukolsky}) is called the tortoise coordinate conveniently. This transformation combines the derivative terms of (\ref{eq:radialTeukolsky}) into a single second-order derivative term
\begin{equation}
	\Delta^{-s} \frac{d}{dr} \left( \Delta^{s+1} \frac{dR}{dr} \right) \rightarrow \frac{d^{2}}{dr_{*}^{2}}Y,
\end{equation}
to generate a Schr\"odinger-like form of radial equation. The transformed radial equation near the event horizon behaves as
\begin{equation}
	\frac{d^{2}}{dr_{*}^{2}}Y + \left(k-is\frac{r_{+}-r_{-}}{2r_{+}(r_{+}+r_{-})}\right)^{2}Y \simeq 0 , \ \ r \rightarrow r_{+},
\end{equation}
where $k \equiv \omega-m\Omega_{+}$, and $\Omega_{+} \equiv \frac{a}{r_{+}^{2}+a^{2}}$ denotes the angular velocity at the event horizon. The asymptotic solution for $Y$ is hence given by
\begin{equation}
	Y \ \simeq \ e^{\pm i (k-i\frac{s}{2}\frac{r_{+}-r_{-}}{r_{+}(r_{+}+r_{-})})r_{*} } \ \sim \ \Delta^{\pm\frac{ s}{2}} e^{\pm i k r_{*}},
\end{equation}
with the expression of $r_{*}$ given in the transformation (\ref{eq:transformationTeukolsky}), and the fact that $\Delta \sim (r-r_{+})$ at the outer horizon. From the definition of $Y$ in (\ref{eq:transformationTeukolsky}), one can subsequently find that the radial part of master function for the Teukolsky equation behaves as
\begin{equation}
	R|_{r\rightarrow r_{+}} \sim
	\left\{
	\begin{array}{cll}
		& e^{ikr_{*}}, & \ \  \textrm{outgoing},\\
		& \Delta^{-s} e^{-ikr_{*}}, & \ \  \textrm{ingoing}.
	\end{array}
	\right.
	\label{eq:behaviorR}
\end{equation}
These two asymptotic solutions behave singular as they oscillate more and more rapidly when the horizon is approaching and expressing the $r_{*}$ by the logarithmic term  in (\ref{eq:transformationTeukolsky}). The ingoing solution also diverges while $s>0$. 

In \cite{Teukolsky:1973ha}, Teukolsky formulate the correct boundary condition at the horizon for the perturbed fields: 
\emph{The perturbed fields, seen by a physically well-behaved observer at the horizon, should be nonspecial. Equivalently, the radial group velocity of a wave packet as measured by a physically well-behaved observer, be negative.} The ``nonspecial" signifies that the field is neither singular nor identically zero. The 4-velocity of the physical observer serves as one of the basis vectors of a local frame, and these basis vectors naturally correspond to a non-singular null tetrad in the NP formalism. This is achieved by choosing the Kerr ingoing coordinates instead of the BL ones and performing a local Lorentz transformation for the null tetrad. As a result, the metric remains non-singular at the horizon, and the well-behaved sector of the NP quantities at the horizon can be identified. Moreover, one can find that if the first condition is satisfied, then the group velocity of the solution is indeed negative. For more details of the analysis, one can refer to  the relevant papers \cite{Teukolsky:1973ha,Teukolsky:1972my,Press:1973zz,Teukolsky:1974yv}. From the above discussion, we can infer that a physically well-behaved observer corresponds to choosing a coordinate system in which the metric is non-singular at the horizon, and the well-behaved sector of the perturbed fields at the horizon can be identified. If this perspective is correct, then it is feasible to find the boundary conditions for perturbations of a general BH.

To construct a nonsingular null tetrad, Teukolsky transform the Kerr metric in BL coordinates $(t,r,\theta,\varphi)$ (singular at the horizon), to the Kerr ingoing coordinates $(v,r,\theta,\tilde{\varphi})$, which are given by
\begin{eqnarray}
	dv &=& dt + dr_{*}, \nonumber \\
	d\tilde{\varphi} &=& d\varphi  + \frac{a}{r^{2}+a^{2}} dr_{*},
	\label{eq:ingoingKerrTransformation}
\end{eqnarray}
and perform a null rotation for the null tetrad. The NP quantity $\Psi$ of Teukolsky equation (\ref{eq:Teukolsky}) transforms as
\begin{equation}
	\Psi =\left[\frac{\Delta}{2(r^{2}+a^{2})} \right]^{-s}\tilde{\Psi},
	\label{eq:nullrotation}
\end{equation}
to eliminate the $\Delta^{-s}$ term of the ingoing solution in (\ref{eq:behaviorR}). The new master function behaves as follow
\begin{equation}
	\tilde{\Psi}(v,r,\theta,\tilde{\varphi})|_{r\rightarrow r_{+}} \sim
	\left\{
	\begin{array}{lll}
		&  e^{-i\omega v + i m \tilde{\varphi} + 2 i k r_{*}} \Delta^{s}, & \\
		&  e^{-i\omega v + i m \tilde{\varphi}}. & 
	\end{array}
	\right.
	\label{eq:behaviorPsi}
\end{equation}
One can immediately verify that the ingoing solution $e^{-i\omega v + i m \tilde{\varphi}}$ is now nonspecial in the new coordinates $(v,r,\theta,\tilde{\varphi})$, which satisfies the condition proposed by Teukolsky. Moreover, one can check that the radial group velocity of this solution $v_g=-d\omega/dk$ is indeed negative, which agrees with the second condition. The outgoing solution still reserves the $r_{*}$ term and hence behaves singular at the horizon. As a result, the correct boundary condition for the radial equation (\ref{eq:radialTeukolsky}) at the event horizon is given by
\begin{eqnarray}
	\Delta^{s} R \sim  e^{-ikr_{*}} &\sim& e^{-i (\omega - \frac{am}{r_{+}^{2}+a^{2}}) \frac{M r_{+}}{\sqrt{M^{2}-a^{2}}} \ln(r-r_{+})} \nonumber \\
	\ \ &=&  e^{-i\frac{(\omega - m \Omega_{+})}{2\kappa_{+}}\ln(r-r_{+})},
	\label{eq:correctBoundaryConditionKerr}
\end{eqnarray}
where $\kappa_{+} \equiv \frac{\sqrt{M^{2}-a^{2}}}{2Mr_{+}}$ represents the surface gravity of the Kerr BH at the event horizon. 

We express the boundary condition in equation (\ref{eq:correctBoundaryConditionKerr}) deliberately in two forms: the first line uses asymptotic quantities ($M$, $a$) for the Kerr BH, while the second line employs quantities ($\kappa_{H}$, $\Omega_{H})$ at the event horizon. The latter appears more concise than the form of the former, because the horizon quantities are more suitable for the boundary condition at the horizon. Moreover, we can derive the latter for a more general rotating BH metric, which leads us to conjecture the bottom expression in (\ref{eq:correctBoundaryConditionKerr}) is universal. This result will be derived in detail in the next subsection.
\subsection{A general metric}
The analysis presented in the previous subsection is limited to the case where the rotating BHs are analytically known and the perturbation equations are separable in coordinates $r$ and $\theta$. However, 
in modified gravity theories, such as the EMS theory \cite{Guo:2023mda,Xiong:2023bpl}, the EsGB theory \cite{Kleihaus:2015aje,Cunha:2019dwb,Berti:2020kgk,Herdeiro:2020wei}, and the dCS theory \cite{Alexander:2009tp,Okounkova:2018pql}, the rotating solutions are no longer of Petrov type D, making the perturbation equations in the NP formalism cannot be separable in $r$ and $\theta$. Moreover, even the KN BH is of Petrov type D, the gravitational and the electromagnetic perturbations cannot be decoupled  \cite{Mark:2014aja}. Furthermore, calculating QNMs in many models involves solving an eigenvalue problem for coupled linear partial differential equations, where the coefficients and rotating BH solutions are often obtained numerically, adding complexity to the analysis of boundary conditions. The analysis in the previous subsection is no longer applicable to such general cases. 

To analyze perturbations in the background of rotating BHs, it is essential to start with a metric that is as general as possible. As noted in \cite{Delaporte:2022acp}, even when the spacetime is constrained to be axisymmetric, stationary, and asymptotically flat, the resulting metric can still be highly complex. However, by imposing the additional condition of circularity, the metric simplifies significantly, reducing the number of free functions to at most four. Circularity \cite{Papapetrou:1966zz} is an isometry of the spacetime which in BL coordinates corresponds to the simultaneous transformation of $t\to-t$ and $\phi_{\rm BL}\to-\phi_{\rm BL}$.  The above general metric that follows from circularity can be written in the Weyl-Lewis-Papapetrou (WLP) form \cite{Papapetrou:1966zz,Kundt:1966zz,Wald:1984rg}
\begin{eqnarray}
	ds^{2} &=& - e^{2\bar{F_{0}}} dt^{2}  + e^{2\bar{F_{1}}} \left( d\rho^{2}+dz^{2} \right) \nonumber \\
	       & & + e^{2\bar{F_{2}}} (d\varphi - W dt )^{2},
	\label{eq:WLPform}
\end{eqnarray}
where the commuting Killing vector fields are $\partial_{t}$ and $\partial_{\varphi}$ and the unknown functions ($\bar{F_{0}},\bar{F_{1}},\bar{F_{2}},W$) depend on ($\rho,z$) only. Some well-known parameterized metrics, such as those proposed in \cite{Cano:2019ore, Johannsen:2013szh, Konoplya:2016jvv}, are fundamentally equivalent to or closely related to specific limits of the WLP metric. Since for axisymmetric and stationary black holes, circularity implies that the angular velocity is constant on the event horizon \cite{Frolov:1998wf},  we can impose a constant-radius event horizon via the following coordinate transformation
\begin{eqnarray}
	\rho &=& \sqrt{r^{2}-r_H r} \sin \theta, \nonumber \\
	z    &=& \left(r-\frac{r_{H}}{2}\right) \cos \theta, 
	\label{eq:CoordinateTransformation}
\end{eqnarray}
	%
where $r_{H}$ represents the horizon radius. Substituting the above transformation into the WLP form (\ref{eq:WLPform}) and redefining the metric functions
\begin{eqnarray}
	e^{2\bar{F_{0}}} &=& e^{2F_{0}} N, \nonumber \\
	e^{2\bar{F_{1}}} &=& e^{2F_{1}} \frac{4(r^{2}-r_{H} r) + r_{H}^{2} \sin^{2} \theta}{4r^{2}}, \nonumber \\
	e^{2\bar{F_{2}}} &=& e^{2F_{2}} r^{2} \sin^{2} \theta,
	\label{eq:redefinition}
\end{eqnarray}
the resulting metric becomes
\begin{eqnarray}
	ds^{2} &=& -e^{2 F_{0}} N dt^{2} + e^{2F_{1}} \left( \frac{dr^{2}}{N}+r^{2} d\theta^{2} \right) \nonumber \\
	& & + e^{2F_{2}} r^{2} \sin^{2}\theta \left(d\varphi-W dt\right)^{2}, 
	\label{eq:ansatz}
\end{eqnarray}
with $N\equiv 1-r_{H}/r$. Note that this ansatz is not written in the standard BL coordinates. In the Kerr limit, the coordinate transformation from the coordinate
system in (\ref{eq:ansatz}) to the  BL coordinates was demonstrated in the Appendix A of \cite{Herdeiro:2015gia}. The Killing vector orthogonal to the horizon is given by $\partial_{t} + \Omega_{H} \partial_{\varphi}$, where 
\begin{equation}\label{angularvel}
	\Omega_{H} \equiv  W|_{r=r_{H}},
\end{equation}
represents the horizon angular velocity. This metric has been employed to calculate the numerical BH solutions in numerous models \cite{Guo:2023mda,Xiong:2023bpl,Cunha:2019dwb,Herdeiro:2020wei,Herdeiro:2014goa,Herdeiro:2015gia,Herdeiro:2015tia,Herdeiro:2016tmi,Santos:2020pmh}, and can hence serve as a paradigm for the following discussion. 

\subsection{Boundary condition at horizon}
In general, the perturbed field can be collectively formulated by a function
\begin{equation}
	\Phi(t,r,\theta,\varphi) = e^{-i\omega t + i m \varphi} \phi(r,\theta), 
	\label{eq:function}
\end{equation}
which can represent one component of the metric perturbations or matter field perturbations (such as the electromagnetic field or the scalar field). 
The spacetime symmetry allows us to naturally separate variables of the perturbation function through the eigenfunctions ($e^{-i\omega t},e^{im\varphi}$) of the Killing vectors ($\partial_{t},\partial_{\varphi}$).
We also assume that the perturbation function does not exhibit a simple power-law behavior as  $(r-r_{H})^{-s}$ ($s$ is a real number) at the event horizon, as this behavior can be factored out through a transformation of the function (e.g., the local Lorentz transformation (\ref{eq:nullrotation})). 
A similar procedure is presented by (27)-(33) in \cite{Blazquez-Salcedo:2023hwg}. We disregard the simple power-law behavior to simplify our discussion. 

In the previous subsection, the boundary condition of perturbations of the Kerr BH at the event horizon formulated by Teukolsky was analyzed within the framework of NP formalism. However, we argue that this formulation can be extended beyond the NP formalism, and the existence of a physically well-behaved observer is the only requirement that needs to be preserved. Since a physically well-behaved observer essentially corresponds to choosing a coordinate system in which the metric is non-singular at the horizon, the first part of the boundary condition formulated by Teukolsky can be reformulated as follows: \emph{the perturbed field at the horizon should be nonspecial and ingoing for a non-singular metric}. The ``ingoing" requirement can be utilized to restrict the form of non-singular metric, as we will see.

The metric (\ref{eq:ansatz}) has a coordinate singularity in the component $g_{rr}$ depicted by the denominator $N$. 
Due to the smoothness of the spacetime manifold at the horizon, there always exists a coordinate transformation
\begin{equation}
	(t,r,\theta,\varphi)   \rightarrow  (v,r,\theta,\tilde{\varphi}),
\end{equation}
that converts the original metric into a non-singular form. 
We require that the tangent vector $\partial_{v}$ and $\partial_{\tilde{\varphi}}$ are the Killing vectors of the spacetime. Consequently, the simplest coordinate transformation can be expressed as 
\begin{eqnarray}
	dv               &=& dt + f(r,\theta) dr, \nonumber \\
	d\tilde{\varphi} &=& d\varphi + g(r,\theta) dr,
	\label{eq:dvdphi}
\end{eqnarray}
which is reminiscent of the Eddington-Finkelstein (EF) type coordinates \cite{Griffiths:2009dfa}. Similar to (\ref{eq:function}), we utilize the eigenfunctions ($e^{-i\omega v},e^{i m \tilde{\varphi}}$) of the new Killing vectors ($\partial_{v},\partial_{\tilde{\varphi}}$) to separate the variables of the perturbation function
\begin{equation}
	\Phi(v,r,\theta,\tilde{\varphi}) = e^{-i\omega v + i m \tilde{\varphi}} \tilde{\phi}(r,\theta).
	\label{eq:behavior}
\end{equation}
According to the above formulation for the boundary condition, the perturbation function is nonspecial at the horizon under a non-singular metric. In other words, the singularity of the original perturbation function at the horizon is attributed to the coordinate singularity of (\ref{eq:ansatz}). Once the coordinate singularity of the metric is removed, the singularity in the perturbation function is also resolved. 

To restrict the perturbation field to be an ingoing wave outside the horizon, the function $f(r,\theta)$ is required to be positive for $r>r_{H}$. From (\ref{eq:behavior}), while we track a fixed phase ($\Theta = -\omega v+m \tilde{\varphi} = \textrm{constant}$) of the perturbation field with $\tilde{\varphi}=$ constant, it can only propagate inward for the radial direction as $t$ increase, because $dv=0,dt>0$. 

Substituting (\ref{eq:dvdphi}) into the metric to eliminate the $g_{rr}$ term which is divergent at $r_{H}$, one can immediately find the simplest form
\begin{equation}
	f(r,\theta) = \frac{e^{F_{1}-F_{0}}}{N},
	\label{eq:f}
\end{equation}
by observing the first terms.
The term $(d\varphi-W dt)^{2}$ in metric (\ref{eq:ansatz}) transforms to $(d\varphi-W dv + \frac{W e^{F_{1}-F_{0}}}{N}dr)^{2}$, which introduces a new coordinate singularity at the event horizon. The subsequent definition of $g(r,\theta)$ is required 
\begin{equation}
	g(r,\theta) = \frac{W e^{F_{1}-F_{0}}}{N},
	\label{eq:g}
\end{equation}
to eliminate such a singularity. The resulting line element 
\begin{eqnarray}
	ds^{2} &=& - e^{2F_{0}}N dv^{2} + 2 e^{F_{0}+F_{1}} dv dr + e^{2F_{1}}r^{2} d\theta^{2} \nonumber \\
	&& + e^{2F_{2}} r^{2} \sin^{2}\theta \left(d \tilde{\varphi} - W dv\right)^{2},
	\label{eq:newansatz}
\end{eqnarray}
has no singularity at the event horizon. Substituting (\ref{eq:f}) and (\ref{eq:g}) into (\ref{eq:behavior}), one can express the perturbation with the original coordinate system $(t,r,\theta,\varphi)$ at the horizon as
\begin{equation}
	\begin{array}{l}
		\Phi(t,r,\theta,\varphi) =  \\ 
		e^{-i\omega t + i m \varphi - i \omega \int \frac{e^{F_{1}-F_{0}}}{N} dr  + i m \int \frac{W e^{F_{1}-F_{0}}}{N} dr } \tilde{\phi}(r,\theta). 
	\end{array}
	\label{eq:integration}
\end{equation}
Recall that $\tilde{\phi}(r,\theta)$ is a nonspecial function. By comparing perturbation function (\ref{eq:function}) with (\ref{eq:integration}), the boundary condition can be reformulated as: At the horizon, the ingoing behavior of the perturbation function can be extracted as an integral form via introducing a new coordinate system, in which the metric is non-singular and possesses an ingoing time coordinate.

These integrals depend on the undetermined metric function $F_{0}$ and $F_{1}$, which requires a specific model. Fortunately, what we only concern is the behavior of the perturbation function at the horizon. The function $e^{F_{1}-F_{0}}$ appearing in (\ref{eq:integration}) is a constant at the horizon, as it is directly related to the surface gravity \cite{Herdeiro:2015gia}
\begin{equation}
	\kappa_{H} = \sqrt{-\frac{1}{2} (\nabla^{a}\xi^{b})(\nabla_{a}\xi_{b})}\ |_{r=r_{H}}= \frac{1}{2r_{H}} e^{F_{0}-F_{1}}|_{r=r_{H}}.
	\label{eq:relation}
\end{equation}
This result is pure geometric, independent of the specific gravity model, and holds for any stationary BH with the Killing field $\xi$ defined by $\xi\equiv \partial_{t} + \Omega_{H} \partial_{\varphi}$ \cite{Wald:1984rg}. 
In other words, the integrand in (\ref{eq:integration}) can be formalized as $\frac{r}{r-r_{H}} [\textrm{constant.} + \mathcal{O}(r-r_{H})]$ near the horizon. This gives the expansion of the integrals at the horizon
\begin{eqnarray}
	\int \frac{e^{F_{1}-F_{0}}}{N} dr &\sim& \frac{1}{2\kappa_{H}} \ln(r-r_{H}) + \mathcal{O}(r-r_{H})^{1}, \nonumber \\
	\int \frac{W e^{F_{1}-F_{0}}}{N} dr &\sim&  \frac{\Omega_{H}}{2\kappa_{H}} \ln(r-r_{H}) + \mathcal{O}(r-r_{H})^{1}. \nonumber \\
	&&
	\label{eq:expansionofintegration} 
\end{eqnarray}
One can immediately find that any practicable modification $\delta f(r,\theta)$ of $f(r,\theta)$ in (\ref{eq:f}) does not affect the leading order of the above expansion (\ref{eq:expansionofintegration}), as $\delta f$ should behave as $\mathcal{O}(r-r_{H})^{0}$ at the horizon due to the regularity requirement of the metric after transformation. The comment on $g(r,\theta)$ is similar. However, the modification can be fixed by requiring that $v$ and $\tilde{\varphi}$ are exactly the EF type coordinates when the Kerr metric is reproduced. In fact, one can check that $v$ defined via (\ref{eq:f}) indeed represents an advanced null coordinate.

The resulting ingoing boundary condition for the perturbation function at the horizon is given by
\begin{eqnarray}
	\Phi(t,r,\theta,\varphi) &\sim& e^{-i\omega t  + i m \varphi - i \frac{(\omega-m \Omega_{H})}{2\kappa_{H}} \ln (r-r_{H}) } \tilde{\phi}(r,\theta), \nonumber \\
	\textrm{i.e.} \ \ \   \phi(r,\theta) &\sim& (r-r_{H})^{-\frac{i(\omega-m\Omega_{H})}{2\kappa_{H}}}  , \ \ r \rightarrow r_{H}.
	\label{eq:expansion}
\end{eqnarray}
The closed form of the boundary behavior (\ref{eq:expansion}) is applicable to any neutral perturbation of the stationary axisymmetric BH that can be described by the metric ansatz (\ref{eq:ansatz}). In above derivation, we only require that the perturbations obey the separation  of variables (\ref{eq:function}) and satisfy the boundary condition formulated by Teukolsky. Moreover, this boundary condition is consistent with (\ref{eq:correctBoundaryConditionKerr}) for the Kerr BH, which is only expressed with quantities at the horizon (surface gravity $\kappa_{H}$ and angular velocity $\Omega_{H}$) instead of the asymptotic conserved charged (such as BH mass and spin) in a simple form. Although there is a potential relation between physical quantities at the horizon and conserved charges, this relation lacks an analytic expression in numerous models where only  numerical solutions exist.


\subsection{Boundary condition at spatial infinity}
The discussion of the boundary condition at spatial infinity for the perturbation function is similar to that at the horizon. One can follow the formulation of the boundary condition previously, and modify the propagation direction of the perturbed field at spatial infinity to be outgoing, since the discussion is based on an isolated BH system without energy injection. Moreover, here we also assume that the perturbation function does not exhibit a simple power-law behavior as  $r^{-s}$ ($s$ is a real number) at the spatial infinity. Following (\ref{eq:behavior}), the outgoing EF type coordinates can be given by
\begin{eqnarray}
	du               &=& dt - f(r,\theta) dr, \nonumber \\
	d\bar{\varphi}   &=& d\varphi - g(r,\theta) dr,
	\label{eq:outgoing}
\end{eqnarray}
where the functions $f(r,\theta)$ and $g(r,\theta)$ are adopted from (\ref{eq:f}) and (\ref{eq:g}), respectively. It is clear that  for fixed $\theta$ and $\bar{\varphi}$, the outgoing radial null geodesics are characterized by $u=\ $constant. 

The perturbation functions in the outgoing EF coordinates become
\begin{equation}
	\begin{array}{l}
		\Phi(u,r,\theta,\bar{\varphi}) = e^{-i\omega u + i m \bar{\varphi}} \tilde{\phi}(r,\theta) = \\  
		\\
		e^{-i\omega t + i m \varphi + i \omega \int \frac{e^{F_{1}-F_{0}}}{N} dr  - i m \int \frac{W e^{F_{1}-F_{0}}}{N} dr } \tilde{\phi}(r,\theta),
	\end{array}
	\label{eq:integrationOutgoing}
\end{equation}
where the only difference with (\ref{eq:integration}) is the opposite sign in front of all integrals. 

The asymptotic  behavior of each metric component  appeared in (\ref{eq:integrationOutgoing}) up to subleading order of $1/r$ is expressed as \cite{Herdeiro:2015gia}
\begin{eqnarray}
	-e^{2F_{0}} N         &\sim& -1+\frac{2M}{r}, \nonumber \\
	\frac{e^{2F_{1}}}{N}  &\sim& 1+\frac{2M}{r}, \nonumber \\
	r^{2}W   			  &\sim& \frac{2 M a}{r},
	\label{eq:AsymptoticBehavior}
\end{eqnarray}	
and the expansion of the integrals in (\ref{eq:integration}) at spatial infinity is given by
\begin{eqnarray}
	\int \frac{e^{F_{1}-F_{0}}}{N} dr   &\sim&  r + 2M\ln r + \mathcal{O}(r)^{-1}, \nonumber \\
	\int \frac{W e^{F_{1}-F_{0}}}{N} dr &\sim&  \mathcal{O}(r)^{-1}.  \label{eq:ExpansionInfinity} 
\end{eqnarray}
Consequently, the boundary condition for the perturbation at spatial infinity is presented as
\begin{equation}
	\phi(r,\theta) \sim e^{i\omega r} r^{2 i M\omega } , \ \ r \rightarrow \infty.
	\label{eq:OutgoingBoundaryCondition}
\end{equation}
This result is consistent with the outgoing behavior of the radial function at spatial infinity presented in \cite{Teukolsky:1973ha}, where the behavior was derived by the asymptotic expansion of the radial function.  

It is worth noting that the boundary conditions (\ref{eq:expansion}) and (\ref{eq:OutgoingBoundaryCondition}) are consistent with those derived in \cite{Chung:2024vaf} for perturbations of the parameterized metric proposed in \cite{Cano:2019ore}, despite the differences in the formulation used to derive them. Since the metric in \cite{Cano:2019ore} was constructed to describe small deviations from the Kerr BH and was expanded to linear order in the coupled parameters, the results presented in this paper are expected to be applicable to more general scenarios. From a broader perspective, we have addressed the singularity of the perturbed fields at the horizon and spatial infinity by separately employing ingoing and outgoing coordinates. However, by adopting hyperboloidal coordinates, the perturbed fields can remain globally regular, with the boundary conditions at both the horizon and spatial infinity naturally satisfied \cite{PanossoMacedo:2024nkw}.

\section{Revisit cases}
\label{section3}
In this section, we verify the main conclusion (\ref{eq:expansion}) for BHs in GR alternative to Kerr and in various gravity theories wherein the boundary condition of perturbations is known. 

\textbf{Other type D BHs.} Besides the Kerr BH, it is known that both the KN BH and the spinning C metric are of Petrov type D. Moreover, the perturbation equations of the two families can be derived  by following the approach of Teukolsky  within the NP formalism \cite{Chandrasekhar:1985,Bini:2008mzd}, although the equation of the KN BH cannot be separable in $r$ and $\theta$. The behavior of perturbations for the KN BH \cite{Dias:2015wqa} and the spinning C metric \cite{Xiong:2023usm,Chen:2024rov}  near the horizon is collectively given by
\begin{equation}
	\phi(r,\theta) \sim \left\{
	\begin{array}{ll}
		(r-r_{H})^{-s-\frac{i(\omega-m\Omega_{H})}{2\kappa_{H}}}, & r\rightarrow r_{H}, \\
		& \\
		e^{i\omega r} r^{2 i M\omega }, & r\rightarrow \infty,
	\end{array}
	\right.
\end{equation}
with an additional spin weight $s$ of the perturbed fields \cite{Xiong:2023usm,Chen:2024rov}. As we mentioned before, the spin weight is irrelevant of the propagation direction of the perturbed fields and can always be eliminated through a redefinition of the perturbed fields. As a result, the boundary condition for perturbed fields  is consistent with (\ref{eq:expansion}).

\textbf{Hairy BHs in EsGB and EMs.} For the static hairy BHs in models beyond GR, such as EsGB and EMs, the calculation of QNMs and the corresponding boundary conditions are presented in \cite{Blazquez-Salcedo:2020rhf,Blazquez-Salcedo:2020caw}. In the static limit, the metric ansatz (\ref{eq:ansatz}) becomes $F_{0}(r,\theta)\rightarrow F_{0}(r)$, $F_{1}(r,\theta)=F_{2}(r,\theta)\rightarrow F_{1}(r)$ and $W=0$. A coordinate transformation $\bar{r} \equiv e^{F_{1}} r$ is necessary to recover the general metric ansatz in the Schwarzschild coordinates
\begin{eqnarray}
	ds^{2} &=& -e^{2F_{0}} N dt^{2} + \frac{e^{2F_{1}}}{N} dr^{2}  + e^{2F_1}r^{2}d\theta^{2}  \nonumber \\
	& &+ e^{2F_1} r^{2} \sin^{2}\theta d\varphi^{2} \nonumber \\
	&=& -A(\bar{r}) dt^{2}+\frac{d\bar{r}^{2}}{B(\bar{r})}+ \bar{r}^{2}d\theta^{2} +\bar{r}^{2} \sin^{2}\theta d\varphi^{2}.\nonumber 
	\label{eq:staticansatz}
\end{eqnarray}
The boundary conditions of perturbations are formulated by \cite{Blazquez-Salcedo:2020rhf,Blazquez-Salcedo:2020caw}
\begin{equation}
	\phi(r)  \sim \left\{
	\begin{array}{ll}
		e^{-i \omega \bar{r}_{*}}, & r\rightarrow r_{H}, \\
		& \\
		e^{i \omega  \bar{r}_{*}}, & r\rightarrow \infty,
	\end{array}
	\right.
	\label{eq:cite}
\end{equation}
with
\begin{equation}
d\bar{r}_{*} = \frac{d\bar{r}}{\sqrt{AB}} = \frac{e^{F_{1}-F_{0}}}{N} dr,
	\label{eq:cite2}
\end{equation}
where 	$\phi(r)$ represents the radial part of the perturbed fields.
This expression is consistent to (\ref{eq:expansion}), and hence the discussion in the previous subsection can be applied to the cases above.

\textbf{A test scalar field.}
When considering the perturbations of a rotating BH, the situation becomes inconclusive. To our knowledge, there has not yet been an established conclusion  for the boundary conditions of perturbations of rotating BHs outside the scope of the NP formalism, except \cite{Chung:2024vaf}. However, we can verify the formula (\ref{eq:expansion}) with several known results. In the Appendix E of \cite{Pierini:2023btw}, Pierini suggested that the tortoise coordinate in the EsGB models, which governs the boundary condition for the full perturbations, can be found by requiring the perturbation equation of a test scalar field to have the Schr\"odinger-like form at the event horizon. This supposition implies a potential assumption that, the boundary behavior of both the gravitational and scalar hair perturbations is identical to that of a simple test scalar field, even though their respective perturbation equations are significantly different. Our derivation aligns with \cite{Pierini:2023btw} so that the boundary condition can be formulated by a single expression (\ref{eq:expansion}).

A test scalar field is described by the Klein-Gordon (KG) equation $\nabla^{\mu}\nabla_{\mu}\Phi(t,r,\theta,\varphi) = 0$, which under the ansatz (\ref{eq:ansatz}) is written as
\begin{eqnarray}
	0 &=&  \left(\frac{r^{3} e^{2 F_1-2 F_0}  \left(\omega -m W\right)^2}{\left(r-r_H\right)}-m^2 \csc ^2\theta  e^{2 F_1-2 F_2}\right) \phi \nonumber \\
	& & +\left[r \left(r-r_H\right) \left(\partial_{r}F_0+\partial_{r}F_2 \right)-r_H+2 r\right] \partial_{r}\phi  \nonumber \\
	& & + \left(\partial_{\theta}F_0+\partial_{\theta}F_2+\cot \theta \right) \partial_{\theta}\phi \nonumber \\
	& & +\partial_{\theta}^{2}\phi +r \left(r-r_H\right) \partial_{r}^{2}\phi,
	\label{eq:testscalar}
\end{eqnarray}
provided that the separation of variables $\Phi(t,r,\theta,\varphi) = e^{-i\omega t + i m\varphi} \phi(r,\theta)$.
There are two approaches to constructing the boundary condition for the above equation. Firstly, one can perform a Frobenius analysis, through replacing $\phi(r,\theta)$ with $(r-r_{H})^{p_{1}} y(\theta)$ or $e^{p_{2}r} r^{p_{3}}y(\theta)$ in equation (\ref{eq:testscalar}) and expanding the equation in Taylor series at the horizon or the spatial infinity, respectively. The exponent ($p_{1},p_{2},p_{3}$) is found at the leading order or the subleading order of the series

\begin{eqnarray}
	p_{1} &=& - i \frac{r_{H}}{e^{F_{0}(r_{H},\theta)-F_{1}(r_{H},\theta)}} \left(\omega- m W(r_{H},\theta)\right) \nonumber \\
	&=& - \frac{i\left(\omega- m\Omega_{H}\right)}{2\kappa_{H}}, \nonumber \\
	p_{2} &=& i \omega, \nonumber \\
	p_{3} &=& 2 i M \omega -1,
\end{eqnarray}
where the second line has used the expressions of the surface gravity (\ref{eq:relation}) and  the angular velocity at the horizon (\ref{angularvel}). Obviously, the above results are consistent with (\ref{eq:expansion}) and (\ref{eq:OutgoingBoundaryCondition}).

The second approach, as described in \cite{Pierini:2023btw}, is formulated by requiring the equation for the test scalar field to have the form
\begin{equation}
	\partial_{r_{*}}^{2} \phi + (\omega-m\Omega_{H})^{2} \phi = \mathcal{O}(r-r_{H}), \ \ r\rightarrow r_{H},
\end{equation}
where the tortoise coordinate $r_{*}$ is defined to eliminate  $\partial_r\phi$ term. The boundary condition at the event horizon is hence given by $\phi|_{r\rightarrow r_{H}} \sim e^{-i(\omega-m\Omega_{H}) r_{*}}$. This is obtained via a replacement $\bar{\phi}(r,\theta)\equiv r e^{(F_{1}+F_{2})/2} \phi(r,\theta)$ in the equation (\ref{eq:testscalar}) and multiply the follow-up equation by $N e^{(F_{1}+F_{2})/2}/r$. It is worthy to emphasize that the term $ r e^{(F_{1}+F_{2})/2}$ is regular at the event horizon and hence does not change the follow-up boundary behavior of the perturbed field. The equation (\ref{eq:testscalar}) then behaves as
\begin{equation}
	\frac{N}{e^{F_1-F_0}} \partial_{r} \left(\frac{N}{e^{F_1-F_0}} \partial_{r} \bar{\phi}\right) 
	+ (\omega-m W)^{2} \bar{\phi} = \mathcal{O}(r-r_{H}),
	\label{eq:testbehavior1}
\end{equation}
at the horizon and 
\begin{equation}
	\frac{N}{e^{F_1-F_0}} \partial_{r} \left(\frac{N}{e^{F_1-F_0}} \partial_{r} \bar{\phi}\right) 
	+ \omega^{2} \bar{\phi} = \mathcal{O}\left(\frac{1}{r}\right),
	\label{eq:testbehavior2}
\end{equation}
at the spatial infinity.
It is evident that the tortoise coordinate can be defined by $dr_{*} \equiv \frac{e^{F_{1}-F_{0}}}{N}dr$ to eliminate the first derivative in (\ref{eq:testbehavior1}) and (\ref{eq:testbehavior2}). The definition of the tortoise coordinate is consistent with the (\ref{eq:f}). Therefore, the boundary conditions at the horizon  (\ref{eq:expansion})  and for at the spatial infinity (\ref{eq:expansionofintegration})  expressed by physical quantities are regained. 


\section{Discussion}
\label{section4}

In this paper, the boundary conditions of neutral perturbed fields at each boundary was derived in the background of a general BH that is stationary, axisymmetric, asymptotically flat, and respects circularity. We first reviewed the boundary condition for perturbations of the Kerr BH derived within the NP formalism and argued that the boundary condition (\ref{eq:correctBoundaryConditionKerr}) expressed with the quantities ($\kappa_{H}$, $\Omega_{H}$) at the event horizon can be generalized to perturbations in a more general BH spacetime. Subsequently, we adopted a commonly used metric (\ref{eq:ansatz}) describing a stationary axisymmetric BH with a constant-radius event horizon. The perturbation function independent of specific perturbed field was presented. The only requirement for this perturbation is that the perturbation observed by a physical well-behaved observer at the horizon is nonspecial and ingoing. The nonspecial perturbed field means neither singular nor identically zero, and the physical well-behaved observer can be indicated by a new metric non-singular at the horizon. We constructed the regular Eddington-Finkelstein type metric (\ref{eq:newansatz}) with coordinates $(v,r,\theta,\tilde{\varphi})$ by eliminating the divergent $g_{rr}$ in (\ref{eq:ansatz}), and expressed the nonspecial perturbation with original coordinates $(t,r,\theta,\varphi)$. We found that the resulting boundary condition (\ref{eq:expansion}) is consistent with (\ref{eq:correctBoundaryConditionKerr}) for Kerr BHs. The boundary condition at spatial infinity was briefly derived as well. We adopted a similar formulation for the boundary condition, with the only difference being the requirement of the outgoing behavior for the perturbed field. 

To verify the reasonability of our results, we revisited the boundary conditions for perturbations of several BH models. These models include KN BHs, the spinning C metric and static hairy BHs in EsGB and EMs theories. For rotating numerical BH solutions, which are the primary focus of this paper, there is no recognized conclusion available for the verification. However, by analyzing the boundary condition at the horizon for a test scalar field, we have validated the correctness of the boundary condition (\ref{eq:expansion}) through two different ways.

This work has extensive applications and potential for further extensions. The boundary condition for perturbations at the horizon can be applied in the study of QNMs in modified gravity theories and in complex environments, and also in calculation of the extreme-mass-ratio-inspiral waveform. Additionally,  the analysis in this paper is based on a general metric that respects circularity, it is interesting to consider the more general case where the circularity is broken \cite{Delaporte:2022acp}, which could lead to the violation of the mirror symmetry of the QNM spectrum \cite{Ghosh:2024het}. Moreover, the  work could be further extended to cases where perturbed fields are charged or massive, which would significantly change the boundary conditions at the horizon or spatial infinity.

\begin{acknowledgments}
	The work is in part supported by NSFC Grant No.12205104, ``the Fundamental Research Funds for the Central Universities'' with Grant No. 2023ZYGXZR079, the Guangzhou Science and Technology Project with Grant No. 2023A04J0651 and the startup funding of South China University of Technology.
\end{acknowledgments}


\end{document}